\documentclass{aa}

\usepackage{url}
\usepackage{graphicx}
\usepackage{natbib}
\usepackage{changebar}

\graphicspath{{fig/}{img/}}

\providecommand{\unit}[1]{\ensuremath{\:\mathrm{#1}}}

\newcommand{\changed}[1]{#1}
\newcommand{\paramset}[1]{$\langle\textrm{#1}\rangle$}

\def\eg{{\it e.g.}\ }

\def\ie{{\it i.e.}\ }

\begin{document}

\titlerunning{Model of coronal energy dissipation based on RMHD}
\title{A simplified numerical model of coronal energy dissipation based
  on reduced MHD}
\author{E. Buchlin\inst{1,2} \and V. Aletti\inst{1} \and
  S. Galtier\inst{1} \and M. Velli\inst{2,3} \and G. Einaudi\inst{3}
  \and J.-C. Vial\inst{1}}

\offprints{E. Buchlin, \protect\url{eric.buchlin@ias.fr}}

\institute{
  Institut d'Astrophysique Spatiale, CNRS -- Universit\'e Paris-Sud,
  B\^at.~121, 91405 Orsay Cedex, France
  \and
  Dipartimento di Astronomia e Scienza dello Spazio, Universit\`a
  di Firenze, 50125 Firenze, Italy
  \and
  Istituto Nazionale Fisica della Materia, Sezione A, Universit\`a di
  Pisa, 56100 Pisa, Italy
}

\date{Received\,:  / Revised date\,:}


\abstract{
  
  A 3D model intermediate between cellular automata (CA) models
  and the reduced magnetohydrodynamic (RMHD) equations is presented to
  simulate solar impulsive events generated along a coronal magnetic
  loop.  The model consists of a set of planes distributed along a
  magnetic loop between which the information propagates through
  Alfv\'en waves.  Statistical properties in terms of power-laws for
  energies and durations of dissipative events are obtained,
  \changed{and their} agreement with X-ray and UV flares observations
  \changed{is discussed}. The existence of observational
  biases is \changed{also} discussed.

  \keywords{MHD -- Sun\,: corona, flares -- turbulence}

}


\maketitle


\section{Introduction}
One of the main unresolved problems in solar physics concerns the
mechanism by which the solar atmosphere is heated from several
thousands degrees in the photosphere to millions in the corona.  It is
now commonly accepted that the ultimate source of energy lies in the
convective motions in and below the photosphere, and that a reliable
model of coronal heating has to deal with the transfer, the storage
and finally the release of this energy into the solar corona.  Several
conceptual models have been proposed, such as Alfv\'en waves, electric
currents and MHD turbulence -- see for a review \citet{zirker} --
where in all cases the magnetic field plays a key role in the
dynamics.

On the other hand, it is also well established that impulsive events
(\eg solar flares, X-ray bright points) are distributed in the corona
over a large range of scales in size, energy and duration
\citep{Dennis85,Crosby93,Pearce93,Krucker98,Aletti00,Asch00b}
and that large events seem to be made up of the superposition of a
myriad of smaller unresolved events. It was \citet{parker88} who
suggested that the active X-ray and UV corona is composed of a swarm
of localized impulsive bursts of energy called nanoflares (or
picoflares).  Much theoretical work has been done to investigate
Parker's conjecture and more generally the statistical nature of the
solar coronal heating.  They mainly follow two complementary schools
which refer on the one hand to the dynamics of complex systems and on
the other hand to fluid mechanics.

The statistical properties of flaring activity allow one to view the
solar corona as a complex system which can be described with cellular
automata (CA) models \citep{Lu91,Lu93,Vlahos95,galsgaard,
George98a,vassiliadis,Isliker00,Isliker01,Charbon,krasno}.  CA models
have become increasingly useful in the study of complex systems
because they permit the study of an entire system without ignoring the
effects of individual components of the system.  There are many
natural applications such as substorms in the magnetotail
\citep{takalo}, star formation in spiral galaxies \citep{lejeune} or
earthquakes \citep{carlson}.  A cellular automaton is based upon the
idea of the locality of influence\,: a system is distributed in space,
and nearby regions have more influence than those far apart (see for
instance \citet{mac2} for a study of a nonlocal communication). A grid
of cells is used to represent the components of a system, and each
cell is given a set of phenomenological rules concerning its
surrounding neighbors. The system evolves over several iterations by
allowing each cell to interact using the given rules. What makes CA so
interesting and useful is that after many iterations they reveal
complex structures and arrangements that form across great distances
even though each cell only takes into account local
information. Self-Organized Criticality (SOC)
\citep{Bak87,Kadanoff89,hwa,sornette} refers to the spontaneous
organization of such \changed{an externally driven system} into a
globally stationary state over many scales.

On the other hand, we find fluid models which give the physical
description that is missing in CA. Much work has been done on
statistical solar flares \citep{Long94,HOO,Einaudi96,
  GalsgaardN,Hendrix96,dmitruk98,Gal98,George98b,Gal99,WG00} but most
of them suffer from the fact that statistical simulations of flares
studied in the context of forced resistive MHD equations are possible
only at the cost of huge computational expenses. Nevertheless it has
been possible to show important properties, \eg that the dissipative
events produced exhibit power-law distributions (for total energy,
peak of luminosity and duration) in agreement with X-ray observations,
but with generally a much smaller ``inertial range'' than the CA
counterpart.

A recent debate about the possible existence of sympathetic flaring,
\ie the correlation in time of two successive events
\citep{Pearce93,Wheatland,Boffetta,Wheatland00,Lepreti,Gal01},
suggests the possibility to dismiss CA as a model of solar flares
since standard CA models do not produce correlated events
(non-Poissonian statistics such as power-law waiting time
distributions). But in fact many CA models exist in the literature
like nonconservative models \citep{chris92} which turn out to be able
to generate the statistics expected for sympathetic flaring. But the
question of the existence of sympathetic flaring in the corona has not
yet found an answer since in particular there is still a debate about
what we mean by event.

The problem of coronal heating is intimately linked to the existence
of nanoflares whose Probability Distribution Function (PDF) in energy
is supposed to be a power-law steeper than that for regular flares.
Let us assume that the PDF in energy $E$ of events is distributed
according to a power-law of index $-\zeta$, \ie $\Pr(E) \propto
E^{-\zeta}$. It is then possible to show that there exists a critical
slope of index $\zeta_c = 2$ \citep{Hudson91}. Indeed, the total
energy released in the corona by events between $E_{min}$ and
$E_{max}$ is $(E_{max}^{2-\zeta} - E_{min}^{2-\zeta}) / (2 - \zeta)$
which means that if $\zeta < 2$ the main contribution comes from high
energy events, whereas if $\zeta > 2$ it comes from smaller events
(the swarm of nanoflares). The average power dissipated in a large
flare is of the same order of magnitude as the total average power
emitted by the corona, $\simeq 10^3 \unit{W}\cdot\unit{m}^{-2}$, which
proves that regular flares can not account for coronal heating since
they are episodic events seen over and above the average coronal
background. It is then natural to think that a swarm of very small and
still unobservable events may dominate the heating process. One of the
main challenges of statistical flare models is to know whether or not
it is possible to produce power-law distributions for any relevant
quantity such as energy, luminosity or duration, and what the
power-law indices are.

\changed{ The aim of this paper is to introduce a hybrid model for a
  solar magnetic loop which is somewhat intermediate between CA models
  and full MHD or reduced MHD models. In this model, we will inject
  and store energy into a coronal loop (our numerical domain) via wave
  propagation from the photosphere (our numerical boundary). The
  trigger for an event is determined in a way analogous to
  conventional CA models, \ie with a threshold in the current.
  However, during the subsequent event the current is dissipated and
  the magnetic field recomputed using Maxwell's equations. Let us note
  that this is a minimal consistency requirement for the field
  evolution which is not always incorporated in CA models. In
  practice, the model allows current concentrations to form
  kinematically (advection from the photosphere), but not dynamically
  (the nonlinear part of the Lorenz force, $\vec{j} \times \vec{b}$).
  The model is non-trivial because of the threshold dynamics of the
  dissipation, which mimics the nonlinear terms, but the model is much
  simpler to integrate than the full MHD equations, therefore allowing
  a fast computation of statistics (events sizes and durations), and a
  comparison both with observations and full numerical simulations.  }

The paper is organized as follows. In Section 2 we give a detailed
description of the CA model and show its basic behavior through some
numerical experiments. In Section 3 the results of a parametric study
are given and discussed. In Section 4 we summarize the properties of
the model, we present a comparison with observations, and we draw some
conclusions.



\section{The model}

In the original 3D lattice model developed by \citet{Lu91} the 
physical quantity defined on each lattice is the magnetic field. 
The system is driven from the outside by adding randomly in space a 
random magnetic field. The process continues until a 
reconnection instability criterion is satisfied at any point of the
3D lattice, \ie until the
magnetic gradient exceeds a critical value at this point.
Then the magnetic field is redistributed (diffused) towards 
neighboring nodes with the possibility to transfer the instability as 
well. The redistribution process stops when the system is completely 
relaxed. Then another random amount of magnetic field is added to the 
system. An event called {\it avalanche} is associated to the rapid
diffusion of the magnetic field.

Subsequent models use the magnetic vector potential $\vec{A}$ rather
than the magnetic field since the divergence-free condition for the
magnetic field is then automatically satisfied. For example in the
recent model developed by \citet{Isliker00,Isliker01} where the 3D
lattice represents an ensemble of magnetic loops, the knowledge of
$\vec{A}$ allows to reconstruct the magnetic field topology and
eventually the structure of the current density. To do so they
introduce the notion of derivative. The present simplified model
belongs to this class of models but only one typical coronal loop will
be considered and simulated. The detailed description of the model is
now given.

\subsection{Basic idea\,: on-off mechanism and turbulence}
\label{subsec:onoff}
\changed{A possible reason for the behavior of the corona is that it
lies in a turbulent state. A model of coronal loops should therefore
allow for the effects of turbulent fluctuations}. This is possible
with CA models at a very superficial level through an on-off
mechanism. The idea behind the threshold dynamics of our model, the
on-off mechanism, is the following. The forcing due to the convective
granules, although applied on a range of scales, has a typical length
scale that is supposed to be far greater than the dissipative
scale. The connection between the forcing and the dissipative length
scales is made through a turbulent mechanism.  During the ``off''
phase, \ie the loading phase, the plasma is in a laminar state where
the dynamics is essentially governed by the linear terms (and the
loading).  Because the system is driven slowly, parts of the system or
even the entire system can stay in principle in this state for very
long periods of time. When sufficient energy is accumulated in the
loop some nonlinear instability appears which triggers the rapid
generation of small scales. The inertial range of the turbulent energy
spectrum extends to small scales and makes the link between the
typical forcing length scale and the dissipative length scale. This
``on'' phase is therefore characterized by a sudden increase of the
dissipative terms in the RMHD equations (see section
\ref{subsec:descr}) leading to a bursty event. Then the system returns
immediately to an ``off'' phase.  The nonlinearities of the RMHD
equations are therefore included in the model in a very schematic way
through a threshold dynamics only.

\subsection{Description of the model}
\label{subsec:descr}

\begin{figure}[htbp]
\includegraphics[width=\linewidth]{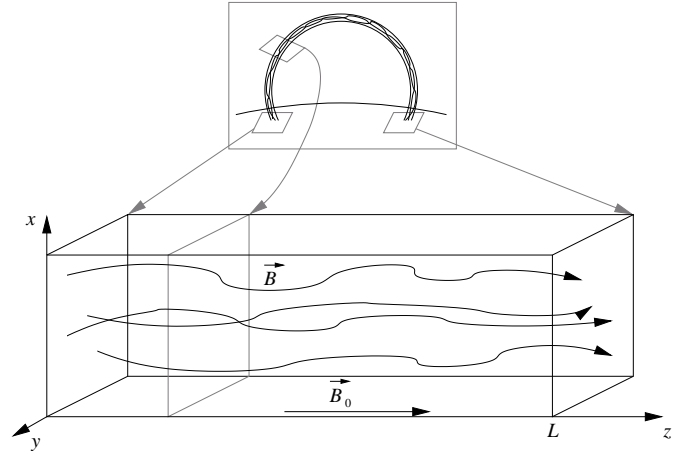}
\caption{\label{fig:gridloop}
  Model of coronal loop used for the simulations\,: the loop is unbent
  into a box, whose both extreme planes represent the photosphere.
  Parallel planes represent slices orthogonal to the local direction
  of the loop.}
\end{figure}

\paragraph{Geometry of the model.}
The  model describes a coronal loop anchored in the photosphere
whose footpoints are randomly moved. The presence of a strong axial
magnetic field leads to essentially 2D dynamics, \ie perpendicular to
the mean magnetic field, for which the approximation of the RMHD
equations \citep{strauss} is a good model. As we can see on
Fig.~\ref{fig:gridloop}, the 3D regular grid is made up of a set of
planes distributed along the loop between which the information
propagates through Alfv\'en waves. Therefore each plane will evolve
essentially independently from each other. Both boundary planes
represent the photospheric footpoints, while the intermediate planes
represent the loop itself, as if it were unbent. The curvature of the
loop is not taken into account since the width of observed coronal
loops (see \eg observations from the TRACE instrument) appears to be
constant along the loop and is much smaller than their length.

\paragraph{RMHD equations.}
Our aim is to compute the temporal evolution of the
velocity field $\vec{v}$ and the magnetic field $\vec{B}$ (or 
$\vec{b} \equiv \vec{B}/ \sqrt{\rho_0\mu_0}$ if we consider only
fields with the same physical dimension). We assume the presence of a
strong and uniform axial magnetic field along the $z$-axis 
$\vec{b}_0 = b_0 \vec{e}_z$ ($\equiv \vec{B}_0/ \sqrt{\rho_0\mu_0}$) 
and we consider small perturbations to this field. We separate
$\vec{b} - \vec{b}_0$ and $\vec{v}$ into parallel components ($b_z
\vec{e}_z$ and $v_z \vec{e}_z$) and orthogonal components
($\vec{b}_\perp$ and $\vec{v}_\perp$). With the following additional
hypotheses, (i) the scales along the $z$-axis are larger than the scales
in the orthogonal directions (gradients along the $z$ axis are
negligible), and (ii) the kinetic pressure is negligible compared to
the magnetic pressure (\ie the plasma is cold, $\beta \ll 1$), we 
then obtain the RMHD equations \citep{strauss}
\begin{eqnarray}
\label{eq:rmhd1}
  \partial_t\vec{v}_\perp
    + (\vec{v}_\perp\cdot\vec\nabla_\perp)\vec{v}_\perp
  & = & 
    b_0\partial_z\vec{b}_\perp
    + \nu\,\Delta_\perp\vec{v}_\perp \\
  \nonumber
  &   &
    + (\vec{b}_\perp\cdot\vec\nabla_\perp)\vec{b}_\perp
    - \vec\nabla(b_\perp^2/2) \, , \\
\label{eq:rmhd2}
  \partial_t\vec{b}_\perp
    + (\vec{v}_\perp\cdot\vec\nabla_\perp)\vec{b}_\perp
  & = &
    b_0\partial_z\vec{v}_\perp
    + \eta\,\Delta_\perp\vec{b}_\perp \\
  \nonumber
  &   &
    + (\vec{b}_\perp\cdot\vec\nabla_\perp)\vec{v}_\perp \, , 
\end{eqnarray}
where $\nu$ and $\eta$ are respectively the kinematic viscosity and 
the magnetic resistivity. To each grid point are associated two scalar 
fields $a^s$, with $s=\pm$, from which the Els\"asser fields are 
derived
\begin{equation}
  \vec{z}^s =\vec{v}_\perp + s \vec{b}_\perp = 
  \vec\nabla_\perp\times a^s \vec{e}_z \, .
\end{equation}
All other fields (magnetic and velocity fields, current density,
vorticity\dots) are derived from $a^s$ by analogy with the standard
magnetic and kinematic equations.

\paragraph{Initial state and boundaries.}
The fields $a^s$ are taken to be zero initially. Each cross-sectional
plane along the loop is periodic for the $a^s$ variables. All
numerical computations for each plane are made in the Fourier space.
We use \changed{inverse Fast Fourier Transforms} (FFT) when we
temporarily need to know the values of a field at some positions in
real space.

\paragraph{Alfv\'en wave propagation.}
In the right hand side of equations
(\ref{eq:rmhd1}-\ref{eq:rmhd2}), the first terms correspond to the
Alfv\'en waves propagation along the $z$-axis\,: $a^+$ propagates to
the bottom of the simulation box (low values of $z$), while $a^-$
propagates to the top. These propagations are modeled by specific
cellular automaton rules used at each time step $\delta t$ of the
simulation\changed{, corresponding to discretization of the Alfv\'en
waves terms of equations (\ref{eq:rmhd1}-\ref{eq:rmhd2}) in the
form\,:
\begin{equation}
  \label{eq:alfven}
  a^s(z, t + \delta t) = a^s(z + s\,\delta z, t)
\end{equation}
}
 There is no loss of energy (or reflection) during
the Alfv\'en wave propagation since the density is assumed to be
constant\,; however we assume that there is a total reflection of the
waves when they reach the photosphere, \ie the two boundary planes of
the simulation box.

\paragraph{Loading.}
The action of the photospheric granules on the magnetic footpoints is
modeled by a random increment $\delta\Psi$ to the fields on both boundary
planes. This increment has random Fourier coefficients but has
globally a power-law energy spectrum in $\sim k^{-\alpha}$
\changed{(the total intensity $P_\mathrm{load} = \int
  (k\,\delta\Psi)^2\,\mathrm{d}\vec{k}$ is also a parameter of the simulation).}
Indeed, observational evidence suggests that the convective layer is
in a turbulent state\,: the photospheric granules exhibit a turbulent
power spectrum of velocity consistent with a Kolmogorov energy
spectrum in $k^{-5/3}$ but only for a narrow inertial range of
wavenumbers ($\ell \sim 1/k < 3 \unit{arcsec}$)
\citep{roudier,chou,espagnet}.  We emphasize that there is no loading
in the other parts of the loop\,: energy is solely carried by Alfv\'en
waves. Furthermore, these waves reflect on both photospheric boundary
planes.

\paragraph{Dissipation criterion.}
We assume that dissipation occurs when an instability criterion is
satisfied, which is the condition that the magnitude of the current
density $\|\vec{J}\|$ exceeds a critical value $J_c$. As
$J_z=\vec{J}\cdot\vec{e}_z$ can be derived from the computed variables
$a^s$ ($J_x$ and $J_y$ are negligible), this dissipation criterion is
likely to have more physical meaning than the criteria used for
example in classical sandpile and in more elaborate models (see
\citet{Charbon} for a review). However, there is still some doubt
about the quality of this criterion, as will be discussed later (see
section \ref{sec:reconnect}).

\paragraph{Reconfiguration of the field.}
When the current density exceeds a given threshold at some real-space
grid points in a given plane, the nonlinear terms of equations
(\ref{eq:rmhd1}-\ref{eq:rmhd2}) which are negligible during the
loading phase (the off phase) become large and dominate the dynamics
of the fields. They are quickly balanced by the dissipative terms when
the energy cascade reaches the dissipative scale. \changed{This}
``on'' phase (see section \ref{subsec:onoff}) is modeled by a
diffusion-like process for the magnetic and velocity fields which
tends to reduce the magnitude of the current density and the
vorticity.

\changed{The detailed algorithm is an updated version of the one
  introduced by \cite{einaudi}. At a time $t$ for plane $z_0$ we
  compute the current density $J_z(x,y,z_0)=-\Delta_\perp A_z$, where
  $A_z = (a^+ - a^-) / 2$ is the magnetic vector potential and
  $\Delta_\perp$ denotes the Laplacian operator in the plane $z_0$.
  If at some grid point $(x, y, z_0)$ the value of $|J_z|$ exceeds the
  threshold $J_c$, $A_z$ is updated in the time $\delta t_c$ (with
  $\delta t_c \ll \delta t$ ) according to the equation $A_z (x, y,
  z_0; t+\delta t_c) = A_z (x, y, z_0; t) - \eta\,\delta t_c\,J_z (x,
  y, z_0; t)$, which corresponds to current dissipation. The current
  density $J_z$ corresponding to $A_z(t+\delta t_c)$ is then computed,
  and this dissipation process is iterated until $J_z$ does not exceed
  the threshold anywhere in the plane $z_0$. However, note that after
  the first iteration of this process, we take $C\, J_c$ as a
  threshold instead of $J_c$. The ``dissipation efficiency'' $C$ is a
  number between $0$ and $1$ which guarantees that the system is in a
  relaxed state after the whole dissipation process.}

\paragraph{Energy release} During this relaxation process, magnetic
and kinetic energies are released. The energy release in each plane
can easily be computed from the variations of $\langle
\vec{b}_\perp^2\rangle$ and $\langle \vec{v}_\perp^2\rangle$ in the
plane. It is the primary variable for our statistics. Note that
topological modifications of the magnetic field may be expected\,: the
connectivity of the magnetic field lines is modified because of the
field diffusion. One of the possible interpretations of this
phenomenon is magnetic reconnection (see however the discussion in
section \ref{sec:reconnect}).

\subsection{Time and space scales.}
\label{sec:scales}

Let $\delta x$ and $\delta z$ be the distance between grid points in
the $x$ (or $y$) and $z$ directions respectively, and let $\delta t$
be the time step. If we assume that the loop length $L$ is $1$ to $100
\unit{Mm}$, then $\delta z$ is $30 \unit{km}$ to $3 \unit{Mm}$ for a
typical resolution of $N_L=30$ points (\ie 30 planes along $z$).  The
analog assumptions for a loop width $\ell$ ($=L/10$) of $0.1$ to $10
\unit{Mm}$ give $\delta x$ between $1.5$ and $150 \unit{km}$ for a
typical resolution of $N_{\ell}=64$.

We can also determine time scales for the model\,: as the Alfv\'en
speed is one in the model units, \ie in units of $\delta z/\delta t$,
we have $\delta t = \delta z / v_A$\changed{\,: the time step is the
time needed by the Alfv\'en wave to propagate from one plane to its
neighbors}. For $B_0 = 10^{-3}$ to $ 10^{-2} \unit{T}$ (\ie $v_A
\approx 1$ to $10 \unit{Mm\,s^{-1}}$ with density $\rho_0 \approx
10^{-12} \unit{kg\,m}^{-3}$), this yields $\delta t$ between $3 \times
10^{-3} \unit{s}$ and $3 \unit{s}$.  Another time scale in the system
is the coherence time of photospheric loading $\delta t_l$. It is
modeled by a periodic re-initialization of the coefficients of the
loading increment $d\Psi$, which occurs every $200$ time steps $\delta
t$, or $0.6$ to $600 \unit{s}$. This is small compared to
observational evaluations of the photospheric coherence time, but the
relevant point is the good separation between time scales\,; besides,
larger values of the photospheric loading coherence time do not alter
the results of the model. When no other indication is given, the time
step $\delta t$ is the unit of time\,; for example, on
Fig.~\ref{fig:initgrowth}, the $x$-axis range maximum is
$10\,000\,\delta t$, \ie between approximately $30\unit{s}$ and
$8\unit{h}$. At last, the shortest time scale in the model is the
cascade time scale $\delta t_c$, which is the time step for
dissipations within a cascade\changed{, and which is analogous to the
non-linear time scale of MHD models}. It is assumed to be completely
separated from the other time scales, \ie $\delta t_c \ll \delta t \ll
\delta t_l$.

\changed{A direct consequence of the separation between the cascade
time scale $\delta t_c$ and the time $\delta t$ of wave propagation
between planes is that the fields of neighboring planes are expected
to be uncorrelated.}



\section{Numerical results of the model}

\subsection{Model behavior}

The simulations presented in this paper have been performed on a local
quadri-RS/6000 IBM workstation at IAS. A typical run of $200\,000$
time steps with a resolution of $N_L \times N_{\ell}^2 = 30 \times
64^2$ takes between 2 days and 2 weeks for one CPU, depending on the
parameters.

\begin{figure}[htbp]
\includegraphics[width=\linewidth]{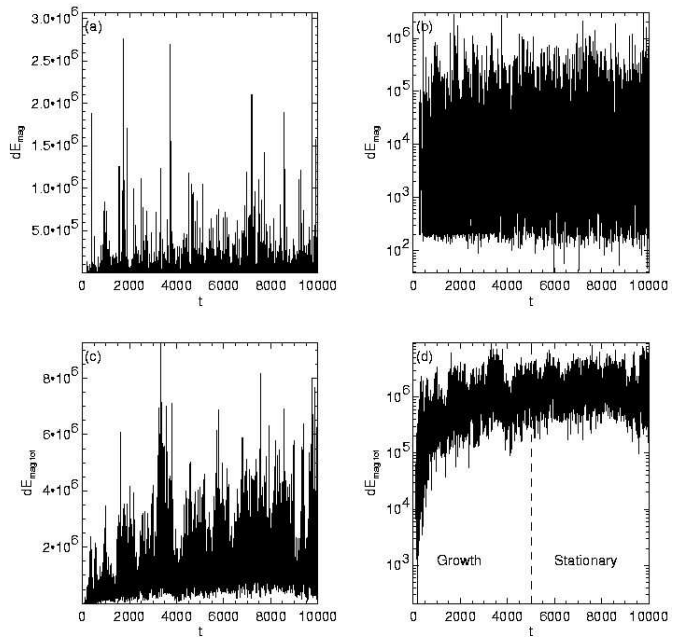}
\caption{\label{fig:initgrowth} Initial growth of energy
  dissipations for parameters \paramset{a} (see Table
  \ref{table:par}): magnetic energies dissipated in the whole
  simulation box (\textbf{a} and \textbf{b}) and in one given plane
  (\textbf{c} and \textbf{d}) are plotted for the 10\,000 first time
  steps of the simulation. \textbf{a)} and \textbf{c)} have linear
  coordinates, \textbf{b)} and \textbf{d)} have semi-logarithmic
  coordinates. Note that no magnetic energy is dissipated in the box
  until $t=90$, and in the plane until $t=240$.  Note also that a
  stationary state is reached from $t \approx 5000$ (bottom right).}
\end{figure}

\paragraph{Initial growths of energy and dissipation.}
As the initial fields in the simulation box are zero, the initial
kinetic and magnetic energies are zero. The loading phase inputs
energy into the system at each time step $\delta t$ which gives a
growth of the total energy of the system as shown on
Fig.~\ref{fig:initgrowth}. Then the current density threshold can be
reached at some points and dissipation occurs, which slows down the
initial energy growth.  At the same time, the average rate of
dissipation increases until a stationary state is reached.

\begin{figure}[htbp]
\includegraphics[width=.85\linewidth]{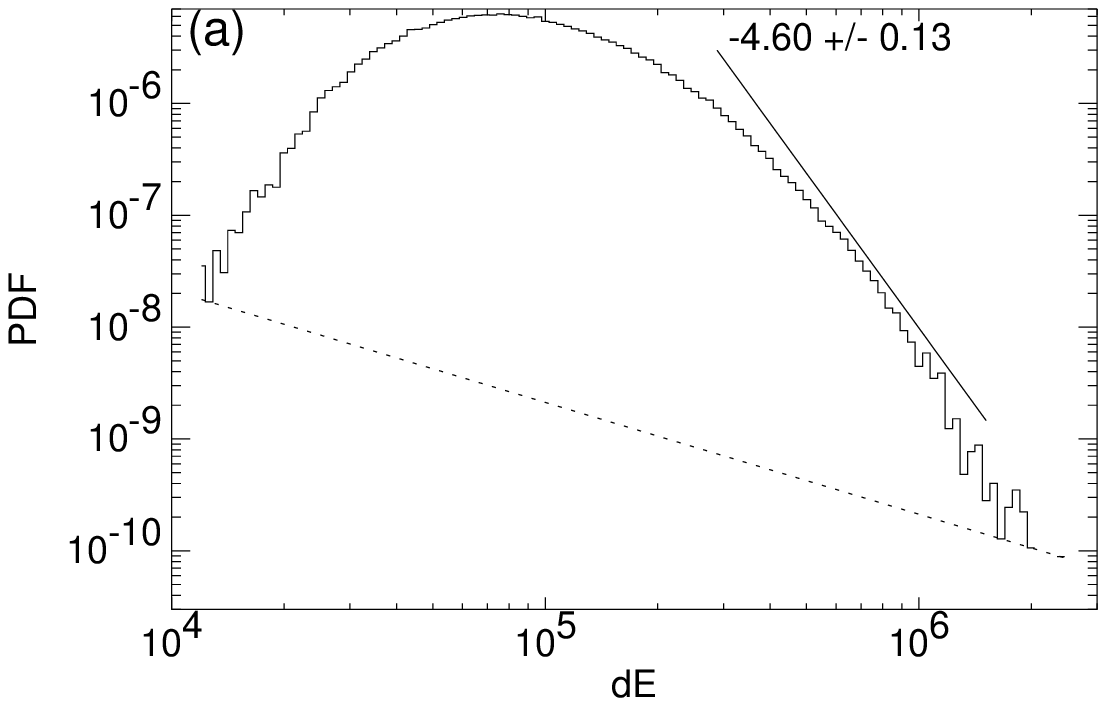}
\includegraphics[width=.14\linewidth]{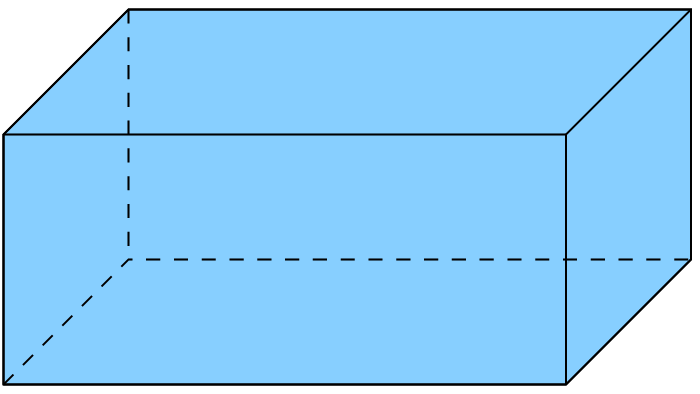}\\
\includegraphics[width=.85\linewidth]{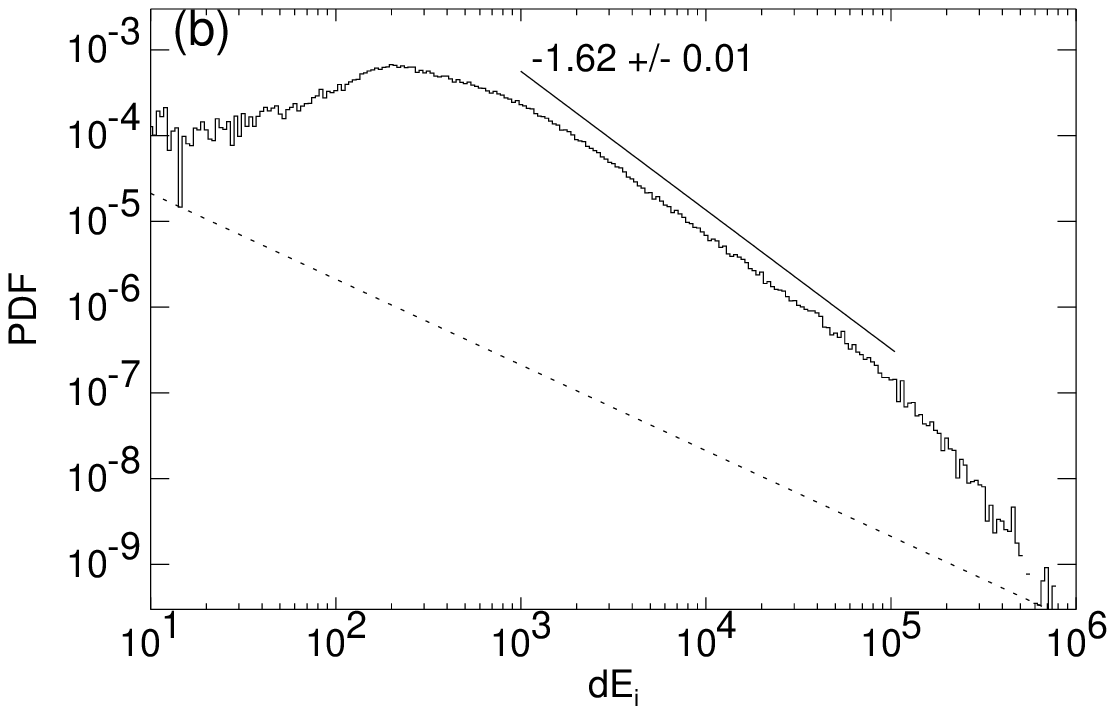}
\includegraphics[width=.14\linewidth]{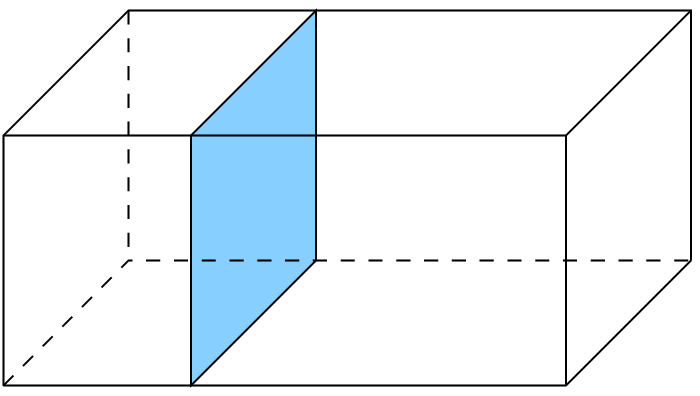}
\caption{\label{fig:refhisto} Histograms of magnetic energy
  dissipations \textbf{a)} in the whole simulation box\,; \textbf{b)}
  in one given plane. The dotted lines correspond to one event per
  histogram bar ($0.02$ decades wide each).  }
\end{figure}

\paragraph{Histograms and fitting methodology.}
The heights of the bars of the histograms we plot are normalized by
their width and they are are divided by the number of events, \ie our
histograms are empirical PDFs. A least-squares linear fit is then
performed in bi-logarithmic axes, on a range determined by visual
inspection of the histogram (see Figure \ref{fig:refhisto}b).  This
gives error bars on the slope of the linear fit, which is the slope of
the expected histogram power-law.  However, we should keep in mind
that the choice of the fitting range often introduces much larger
error bars (typically $\pm 0.1$ to $0.2$) than the error bars of the
least-squares linear fit of the slope (typically $\pm 0.01$ to
$0.05$). The error bars we give from now are conservative
estimates taking into account the fitting range uncertainty.

\paragraph{Choice of the variable used for the statistics and general
shape of the PDF.}

Former studies (\citet{Aletti00, Aletti01}) plotted the
histograms or PDFs of the magnetic energy dissipation $\Delta
E_\textrm{\scriptsize tot}$ calculated in the whole simulation box
(Fig.~\ref{fig:refhisto}).  The global shape of the PDF was a
Gaussian.  A power-law $\Pr(\Delta E_\textrm{\scriptsize tot}) \propto
\Delta E_\textrm{\scriptsize tot} ^ {-6}$ seemed to appear as
a deviation from the Gaussianity in the tail of the distribution, but
it only spanned half a decade, which makes it perhaps not so
relevant. 
On the contrary, we choose to plot the PDF of the magnetic energy
$\Delta E_i$ dissipated in {\em one} given plane $i \in [1,N_L]$
(Fig.~\ref{fig:refhisto}b). As the computations are done in the Fourier
space, this is our primary variable. The power-law that can be fitted
to the PDF of this variable has a much wider range (more than $2$
decades) and is much less steep (the index is between $1$ and $2$)
than in the former case.

This can be explained by a Central Limit Theorem after remarking that
$\Delta E_\textrm{\scriptsize tot} = \sum_i \Delta E_i$ \changed{and
  that the $E_i$'s are quasi-independent (as expected, the correlation
  between fields in neighboring planes is very low)}, thus the PDF of
$\Delta E_\textrm{\scriptsize tot}$ is the convolution of the PDFs of
all $\Delta E_i$ for $i \in [1,N_L]$. The difference between the PDFs
in both cases stresses the importance of the choice of the variable
used for the statistics.  It also emphasizes that in the case of the
statistics of observational data, we have to be careful about the
definition of an ``event''.


\changed{Both distributions of $\Delta E_i$ and of $\Delta E$ show a
  maximum. In the case of $\Delta E_i$, it is a consequence of the
  finite range of the power-law distribution\,; the position of the
  maximum depends the average event size and on the slope. In the case
  of $\Delta E$, knowing the distributions $\Delta E_i$ in all planes
  $i$, it can be seen as a simple consequence of the Central Limit
  Theorem.}

\begin{figure}[htbp]
\includegraphics[width=\linewidth]{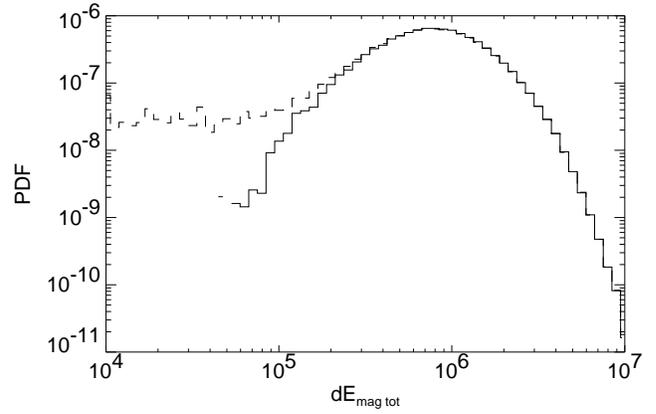}
\caption{\label{fig:inithisto} Effect of initial energy growth on the
  statistics of magnetic energy dissipations in the entire box\,: one
  histogram (solid line) only takes into account stationary state
  events, whereas the other one (dashed line) also takes into
  account the events produced during the initial phase.  }
\end{figure}

\paragraph{Effect of initial growth on statistics.}
During the initial energy growth, the PDF of the energy of events is
different than during the stationary state. In particular, it is
shifted to the left, \ie the events are smaller. As a result, the left
part of the PDF of events energy gets higher than what it would be if
stationary state events only were taken into account, as seen on
Fig.~\ref{fig:inithisto}. As we are interested in stationary state
events, we choose to exclude events occuring during the initial energy
growth from the statistics.

\begin{figure}[htbp]
\hspace*{-8mm}
\includegraphics[width=1.1\linewidth]{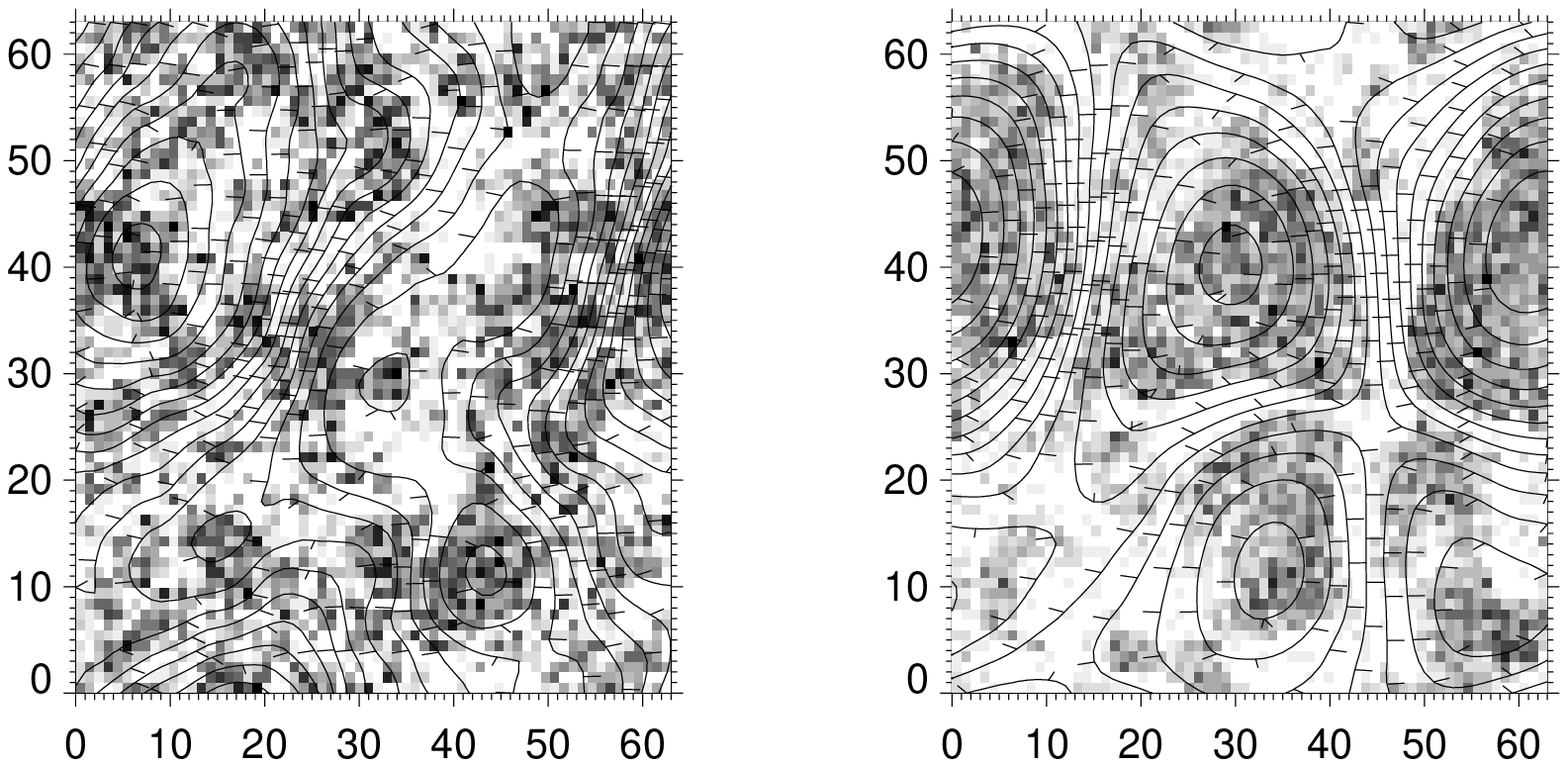}\\
\hspace*{-4mm}
\includegraphics[width=.5\linewidth,height=.5\linewidth]{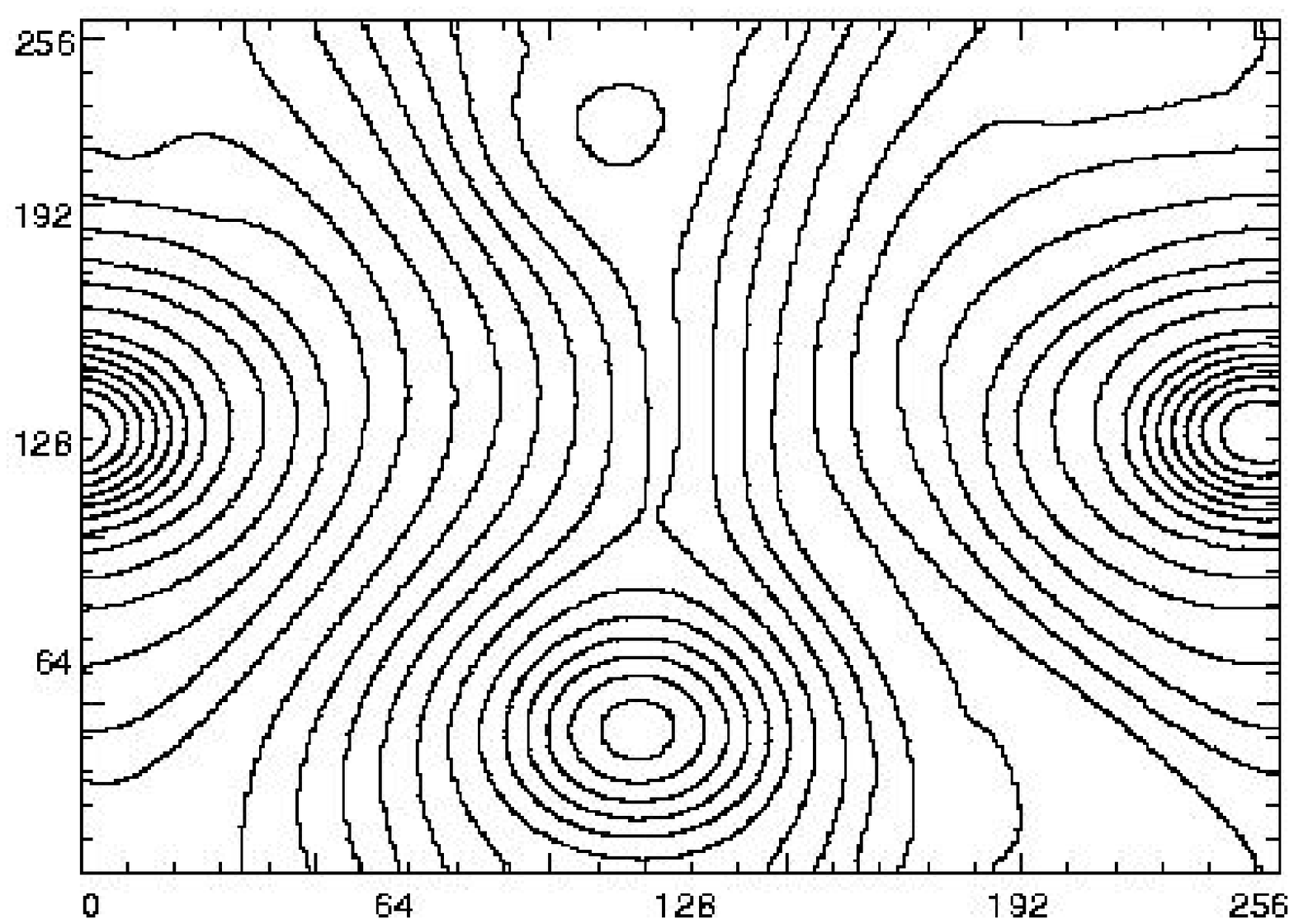}
\hfill
\includegraphics[width=.5\linewidth,height=.5\linewidth]{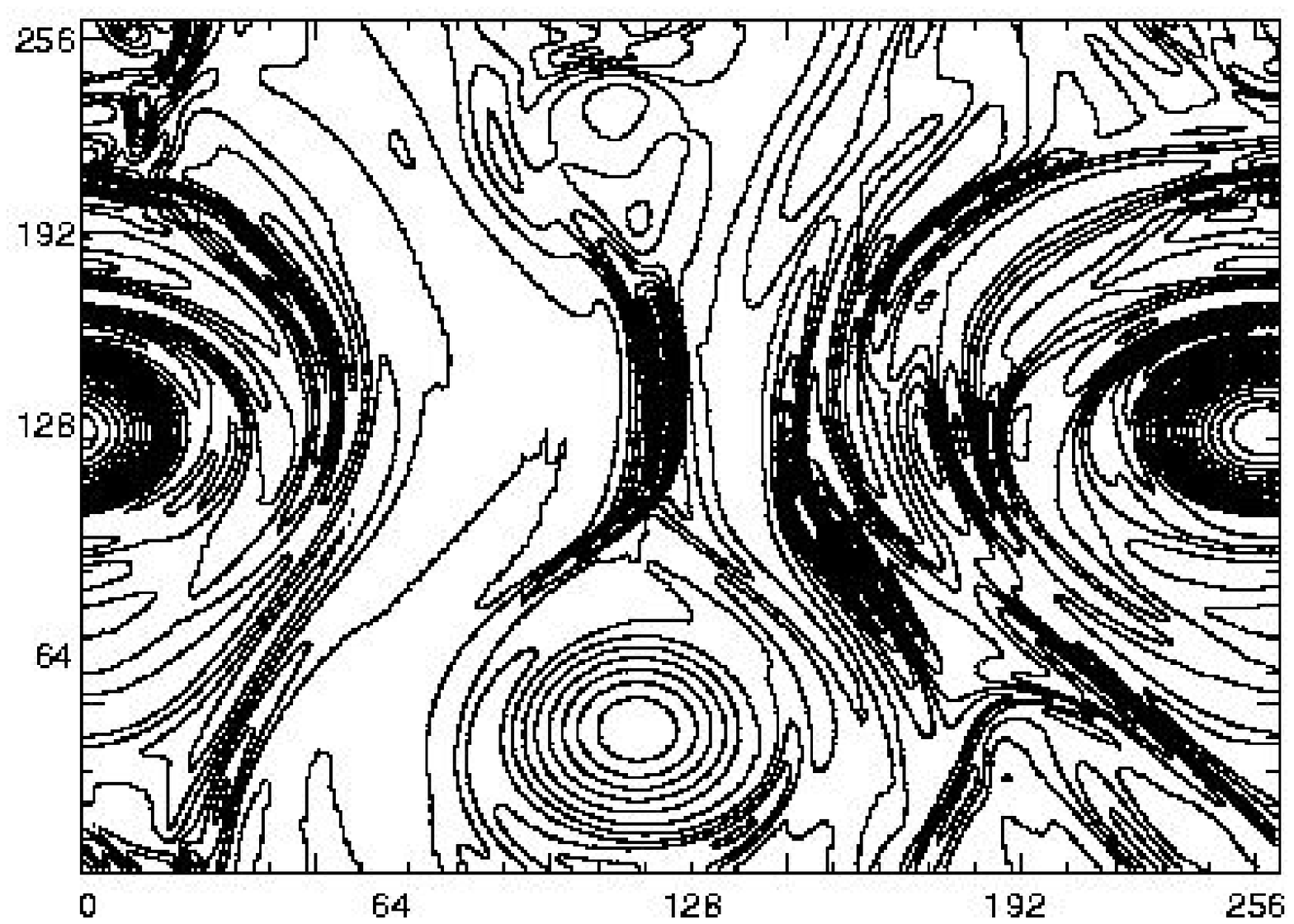}
\caption{\label{fig:typfields} Top\,: magnetic and current density fields
  produced in a plane of the simulation box, for parameters sets
  \paramset{i} ($\alpha=2$, left) and \paramset{o} ($\alpha=4$,
  right). On both samples, the magnetic field lines are superimposed
  on a grayscale map of $J_z^2$ with large values in black. Bottom:
  magnetic field (left) and current density contours (right) issued
  from numerical integration of RMHD equations (courtesy
  \citet{George98b}).}
\end{figure}

\paragraph{Typical fields.}
As the model is built on phenomenological evolution rules, it is not
expected that the fields produced by the simulation coincide with any
real picture. However, as we have tried to be as close as possible to
the original MHD equations it is interesting to see how far the fields
are realistic and what the limits of this phenomenological model could
be. Typical magnetic and current density fields for $\alpha=2$ and
$\alpha=4$ are shown on Fig.~\ref{fig:typfields}. On both samples but
especially for high values of $\alpha$, we can notice that high
current densities occur in magnetic ``islands'' and in regions where
magnetic field densities are high. We do not observe many
structures such as current sheets or possible reconnection sites, as
will be discussed in section \ref{sec:reconnect}, although they are
more present for small values of $\alpha$. Large-scale photospheric
forcing (large $\alpha$) leads to large-scale structures.

\subsection{Parametric study of event energy PDFs}

A parametric study is performed in order to explore the influence of
the simulation parameters on the magnetic energy dissipations PDFs.  A
reference set of parameters, called \paramset{a}, is chosen (see Table
\ref{table:par}), and it gives the PDF shown on
Fig.~\ref{fig:refhisto} . The PDFs obtained for other sets of
parameters will be compared to the PDF obtained for \paramset{a}. Most of the
sets of parameters correspond to the modification with respect
to \paramset{a} of
one parameter (dissipation efficiency $C$, magnetic resistivity
$\eta$, loading spectrum index), which is in italic in Table
\ref{table:par}. All simulations were performed on $200\,000$
timesteps, which seems sufficient to achieve a long stationary state
after the initial energy growth phase, and to achieve good statistics
for the PDFs. Histograms were done with data from the $100\,000$ last
timesteps.

Other parameters, which are not changed during the study, include the
grid size (see above) and the current density
threshold $J_c = 300$. A higher grid resolution would have been
interesting so as to get a broader wavelength range, but it would
need a rescaling of other parameters and longer computation times. The
current density threshold fixes the scale of current density, so its
value has no intrinsic meaning.

\begin{table}
\centering
\caption{\label{table:par}Sets of parameters used for the parametric
  study. $C$ is the dissipation efficiency, $\eta$ is the magnetic
  resistivity and $\alpha$ is the index of the 1D power-law loading
  spectrum. Parameters which are different from parameters set \paramset{a}
  are shown in italic font.
  }
\begin{tabular}{clll}
  \hline\noalign{\smallskip}\hline\noalign{\smallskip}
    & $C$  &  $\eta$           &  $\alpha$ \\
\noalign{\smallskip}\hline\noalign{\smallskip}
\paramset{a} & $0.5$ & $1\cdot 10^{-3}$ & $ 5/ 3$ \\

\paramset{b} & $\mathit{0.1}$ & $1\cdot 10^{-3}$ & $ 5/ 3$ \\

\paramset{c} & $\mathit{0.3}$ & $1\cdot 10^{-3}$ & $ 5/ 3$ \\

\paramset{d} & $\mathit{0.7}$ & $1\cdot 10^{-3}$ & $ 5/ 3$ \\

\paramset{e} & $\mathit{0.9}$ & $1\cdot 10^{-3}$ & $ 5/ 3$ \\

\paramset{f} & $0.5$ & $\mathit{3\cdot 10^{-4}}$ & $ 5/ 3$ \\

\paramset{g} & $0.5$ & $\mathit{3\cdot 10^{-3}}$ & $ 5/ 3$ \\

\paramset{h} & $0.5$ & $1\cdot 10^{-3}$ & $\mathit{ 3/ 2}$ \\

\paramset{i} & $0.5$ & $1\cdot 10^{-3}$ & $\mathit{ 2   }$ \\

\paramset{j} & $0.5$ & $1\cdot 10^{-3}$ & $\mathit{ 7/ 3}$ \\

\paramset{k} & $0.5$ & $1\cdot 10^{-3}$ & $\mathit{ 8/ 3}$ \\

\paramset{l} & $0.5$ & $1\cdot 10^{-3}$ & $\mathit{ 3   }$ \\

\paramset{m} & $0.5$ & $1\cdot 10^{-3}$ & $\mathit{10/ 3}$ \\

\paramset{n} & $0.5$ & $1\cdot 10^{-3}$ & $\mathit{11/ 3}$ \\

\paramset{o} & $0.5$ & $1\cdot 10^{-3}$ & $\mathit{ 4   }$ \\ 
\noalign{\smallskip}\hline

\end{tabular}
\end{table}

\paragraph{Dissipation efficiency.}
Dissipation efficiency tells how much the system gets relaxed after a
series of iterative dissipations\,: the current density threshold $J_c$
is replaced by a new threshold $C\cdot J_c$ after the first
dissipation. With respect to the value $C = 0.5$ used in parameters set
\paramset{a}, the dissipation efficiency can be set to almost any
value of its range $[0, 1]$ of valid values with almost no visible
change in the PDFs.

\paragraph{Magnetic resistivity.}
Magnetic resistivity $\eta$ could vary in the range $[3\cdot 10^{-4},
3\cdot 10^{-3}]$. A numerical stability analysis, given the time step
fixed to one and the wavenumber range, shows that higher values of
$\eta$ would result in numerical instability. On the other hand, lower
values of $\eta$ would lead to longer computational time.  However,
$\eta$ has mainly an influence only on the dissipation process length;
a change in the value of $\eta$ has little influence on the histograms
of dissipated energies.

\begin{figure}[htbp]
\includegraphics[width=\linewidth]{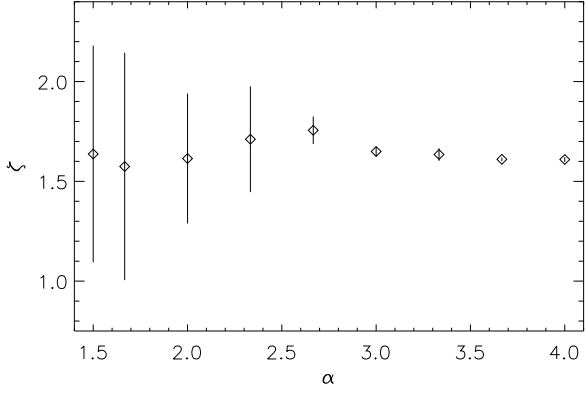}
\caption{\label{fig:powind}
  Event energy PDF power-law index $\zeta$ as a function of the loading
  energy spectrum index $\alpha$ (Table \ref{table:powind}). On the
  left, the big error bars are a consequence of the fact that the
  fitting range is not well defined.}
\end{figure}

\paragraph{Loading spectrum.}
The reference parameters set \paramset{a} has a loading spectrum index
$\alpha=5/3$, corresponding to the spectrum in the inertial range of
Kolmogorov turbulence. In parameters sets \paramset{a} and
\paramset{h} to \paramset{o}, $\alpha$ varies from $3/2$ to $4$ by a
maximal interval of $1/3$. Another series of simulations was performed
with lower loading power values \changed{to explore the influence of
the ratio $P_\mathrm{load} / J_c$. For both series, for high values of
$\alpha$, the power-law is well defined (3 to 4 orders of magnitude
wide). Its slope index is approximately $-1.6$, and this value depends
neither on the loading spectrum index $\alpha$ (as seen in Table
\ref{table:powind} and Figure \ref{fig:powind}, and as will be
discussed in section \ref{sec:reconnect}) nor on the loading intensity
$P_\mathrm{load}$}. For low values of $\alpha$, however, power-laws
were difficult to obtain, and their slopes were sensitive to both
loading spectrum index and loading intensity.

\begin{table}[htbp]
\centering
\caption{
  \label{table:powind}
  Variability of the event energy PDF power-law index $\zeta$ as a
  function of the loading spectrum index $\alpha$.
  }
\begin{tabular}{cll}
  \hline\noalign{\smallskip}\hline\noalign{\smallskip}
    & $\alpha$ & $\zeta$ \\
\noalign{\smallskip}\hline\noalign{\smallskip}
\paramset{h} & 3/2     & $1.64 \pm 0.54$ \\
\paramset{a} & 5/3     & $1.57 \pm 0.57$ \\
\paramset{i} & 2       & $1.61 \pm 0.33$ \\
\paramset{j} & 7/3     & $1.71 \pm 0.27$ \\
\paramset{k} & 8/3     & $1.76 \pm 0.07$ \\
\paramset{l} & 3       & $1.65 \pm 0.02$ \\
\paramset{m} &10/3     & $1.63 \pm 0.03$ \\
\paramset{n} &11/3     & $1.61 \pm 0.01$ \\
\paramset{o} & 4       & $1.61 \pm 0.01$ \\
\noalign{\smallskip}\hline
\end{tabular}
\end{table}

\subsection{Statistics of durations of events}

\label{sec:duration}

\begin{figure}[htbp]
  \includegraphics[width=\linewidth]{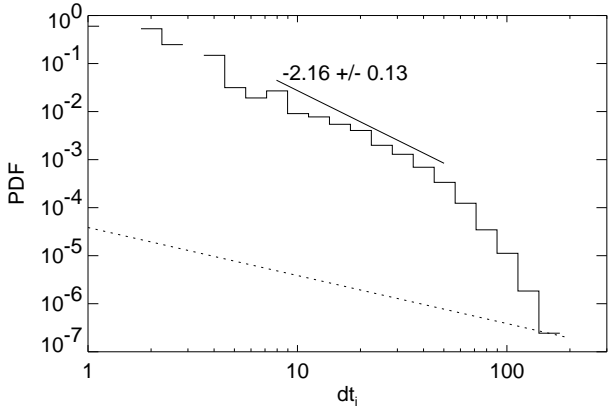}
  \caption{\label{fig:histot}Histogram of events durations for
    parameters set \paramset{l}.}
\end{figure}

Durations of events extend over two decades. They are indeed a
discrete variable, which is a multiple of the cascade time step
$\delta t_c$, and their maximum value is a few
hundreds times $\delta t_c$. Histograms can be obtained, although their
width is too narrow to perform relevant power-law fitting
(see figure \ref{fig:histot}).

\begin{figure}[htbp]
  \includegraphics[width=\linewidth]{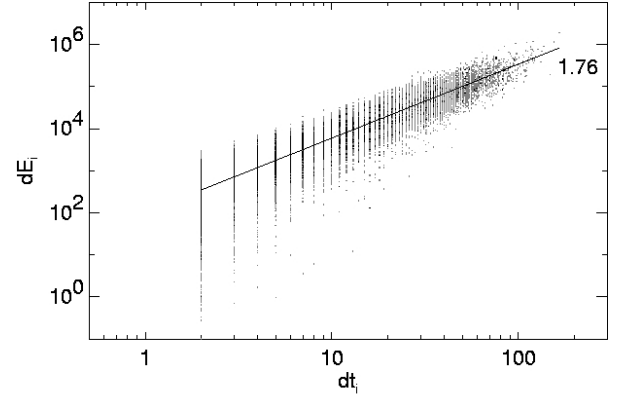}
  \caption{\label{fig:etscat}Correlation between events duration and energy for
    parameters set \paramset{l}.}
\end{figure}

\begin{figure}[htbp]
  \includegraphics[width=\linewidth]{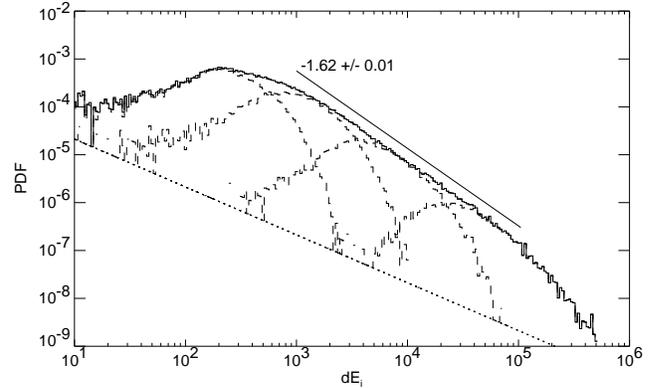}
  \caption{\label{fig:popt}Event energy histograms for different event
    durations ranges, which form a partition of the whole duration
    range, from low duration (left, dashed) to long duration (right,
    dashed). The sum of these histograms (\ie the event energy
    histogram of all events as in figure \ref{fig:refhisto}b) is
    shown as a solid line.}
\end{figure}

The duration of an event is correlated with its energy, like $dE_i
\propto dt_i^{1.76}$, as seen on the scatter-plot on figure
\ref{fig:etscat}. Another way to visualize this correlation is to
select events according to their duration, and to plot the histogram
of energies of events from this population, as shown on figure
\ref{fig:popt}. One possible observational consequence could be that
missing long-duration events, due for example to short observation
times, can lead to energy histograms with narrower ranges and steeper
slopes.

\section{Discussion and Conclusions}

\subsection{Properties of the model}

The CA model presented here differs from previously presented work in
two features. On the one hand, the energy pumping due to photospheric
motion is known quite accurately via
solution of the Alfv\'en wave propagation equation with reflecting
boundary conditions. Each cell therefore receives and sends energy to
neighboring cells along the loop axis via a wave equation. Energy
redistribution to cells on the same loop plane however occurs using an
instability and redistribution criterion. This criterion is a
threshold in current, and satisfies the basic requirements for
magnetic fields of divergence-free conditions and realistic current
redistribution.

\paragraph{Power-law slope of energy distributions as a function of
parameters.}

When parameters sets lead to wide and robust power-law statistics of
events energies, it seems that the slope of these power-laws is quite
the same and takes a value close to $-\zeta=-1.6$. If this behavior is
confirmed by further parametric studies, it could be interpreted as
the ``universal'' behavior of SOC systems, described for example in
\cite{Bak87}.

It is however interesting to note that \citet{George98a} observe in
their model a variability of the slope of event energy (and also peak
luminosity and duration) as a function of the loading\,: the slope is
steeper when the amplitude distribution of the loading increments has
a steeper power-law. However, this discrepancy is due to the
differences in the way the system is driven\,: their loading
increments are discrete in space whereas our loading has a varying
spatial spectrum and no power-law amplitude distribution variation.

\paragraph{Reconnection, and the quality of the dissipation criterion.}
\label{sec:reconnect}
Snapshots of the superposition of the magnetic lines with the current
densities (see figure \ref{fig:typfields}) reveal that the sites of
dissipation often do not correspond to reconnection of magnetic field
lines\,; the reason is that one may have intense currents (and our
dissipation criterion is based precisely on current intensity) without
having a topology of the field where reconnection occurs
(instabilities such as resistive kinks and/or tearing modes are
triggered by a combination of currents and current gradients).  Though
the typical field structures observed resemble fields from turbulence
simulations, one sees that our simplified model distributes
dissipation in a different, more homogeneous way.  Of course, a
cellular automaton model, which models the non-linear terms of the
equations through simple threshold dynamics, is not supposed to
generate such structures, but we can try to understand what can be
done to improve the model. The physical quantities available in our
model make it for example possible to use a more elaborate dissipation
criterion which would model more accurately the reconnection
instability threshold, such as for example introducing a combined
criterion on current and current gradients as a trigger for
relaxation.

\subsection{Comparison with observations}

Impulsive coronal events are statistically distributed over an energy
range of some eight orders of magnitude. Since Parker's idea of the
existence of nanoflares, it has been thought that at some point in the
quest of small scale coronal structures we will break the
self-similarity of the solar corona by the observation of a steepening
of the flare distribution with finally a power-law index $\zeta$
greater than the critical value of $2$. Recent data analyses
\citep{Krucker98,Asch00b,parnell00} seem to show this behavior with
observed values of $\zeta$ going up to $2.6$. Empirical formulas have
been used here to determine flares energies from observed luminosities
but new analyses of the data \citep{Asch02,Benz} reveal in fact the
existence of a bias due to the finite range of temperature on which
the observations are made. The correction of this temperature bias
leads eventually to a value of $\zeta$ close to $1.6$ \changed{valid
  for the whole range of energies, from unresolved, X-ray observations
  with the Solar Maximum Mission (SMM) to Extreme Ultra-Violet
  observations with the Extreme-Ultraviolet Imaging Telescope on the
  Solar and Heliospheric Observatory (SoHO/EIT) and the Transition
  Region And Coronal Explorer (TRACE) -- see \cite{Asch00b}. Such an
  observational bias, as well as the one described in the end of
  paragraph \ref{sec:duration}, show the importance of defining well
  what an event is and what its characteristics are, when observing
  the corona but also when using statistical flare models.}

\changed{The comparison between model distributions and observed
  distributions is actually not an easy task, even if it is tempting
  to compare the power-law slope $1.6$ obtained by our model to the
  observed global slope of $1.6$.  The first pitfall for comparison
  between statistical results of observations and models may be linked
  to the spatial resolution of observations ($\approx 100 \unit{km}$
  at best) compared to the dissipative scales of the system ($\approx
  100 \unit{m}$).} It is clearly shown in this paper that the plot of
the PDF of the magnetic energy dissipated in a given plane of the
model has a shape quite different from the PDF of the same observable
but for the whole simulation box. The bias here consists of a
steepening of the power-law slope with an index greater than $2$. This
result suggests that the limited instrumental resolution may be a
source of error as well. The SOHO/EIT observations analyzed by
\citet{Aletti00} seem to illustrate quite well this interpretation\,:
the pixel intensity distribution power-laws are steeper (with
indices going up to $5.6$) at lower resolution. Most of the coronal
structures in the quiet Sun are indeed smaller than the spatial
resolution of EIT.  Besides, the domain where the power-law is fitted
on the PDF is reduced which leads to larger error on the value of the
index.  Reliable statistical results could be accessible with higher
instrumental resolution but also by using mathematical tools like for
example Pearson's method \citep{Podla02}.

\changed{However, we think that resolution has not always such a
  dramatic effect on the slope of the PDFs, and that it is still
  interesting to model statistics of individual events (\ie in one
  plane, in the case of our model). The convergence to a Gaussian when
  summing the energies of events before doing statistics, predicted by
  an argument lying on the Central Limit Theorem, may indeed be much
  slower in the case of observed, real micro-events than in the ideal
  case of independent events with low-moments distributions: the
  distributions of real events energies are much wider than the
  modeled distributions and their moments may be greater. As a result,
  depending on the observational conditions, there could still exist a
  quite wide power-law after summation (due to lack of resolution) of
  a large but not too large number of events, and the slope of this
  distribution may be still close to the slope of the original
  distribution of individual events. In this context the slope $\zeta
  = 1.6$ we obtain is rather encouraging.}

Another prediction of our model is that durations and energies of
events scale like $dE_i \propto dt_i^{1.76}$, or, equivalently, $dt_i
\propto dE_i^{1/1.76}=dE_i^{0.57}$. This exponent $0.57$ is in quite
good agreement with \citet{Berghmans}, who report that observed events
durations scale like their radiative loss at the power $0.5$.

We should however emphasize that statistical flare models usually give
energy dissipations, while observations give luminosities at some
given wavelengths and the infered energies depend on models.
It is therefore crucial in the future to develop models including the
production of \emph{observable} quantities, in order to provide
stronger links between models and observations but also to quantify
more precisely the weight of observational biases.  The first
agreements obtained during the last decade between statistical
predictions made by theoretical models and observations are however
very promising.

\subsection{Summary and conclusion}

In this paper we have presented a three-dimensional simplified model
inspired by the RMHD equations whose first version was introduced by
\citet{einaudi}. This model mimics a coronal magnetic loop anchored in
the photosphere whose footpoints are driven randomly by convective
motions.  The slow driving of the magnetic footpoints leads to storage
of energy along the coronal loop and eventually to dissipation through
impulsive events. The characteristics of the model are the
following\,: (i) the model describes Alfv\'en wave propagation along a
loop exactly\,; the internal structure of the loop is described by a
set of planes distributed along the axis and ending in the photosphere
from which the information propagates\,; (ii) the external forcing
applied to the two boundary planes is expressed as a turbulent
spectrum in Fourier space\,; (iii) when the criterion of instability
is satisfied, the dissipation of current density and vorticity takes
place non-locally in Fourier space but still locally in physical
space\,; (iv) Fast Fourier Transforms (FFT) are implemented in the
numerical code to use the dual physical/Fourier space.

A numerical study has allowed to quantify the role of the parameters,
especially forcing, in the behavior of this model and in statistical
properties of coronal events. The slope of event energy histograms was
found to be almost constant, in accordance with a SOC-like
``universal'' behavior, and consistent with the values given by
observations. Event durations statistics were performed, and
correlations with event energies are also compatible with
observations. Different possible observational biases were pointed
out, all of them resulting in a narrower power-law range on histograms
and in a steeper slope than in the statistics of all elementary
events\,: a bias due to the limited spatial resolution, which gives a
possible interpretation of recent observations made with the
instrument EIT on board SoHO, and a bias due to limited observation
durations.

Improvements are always possible to provide a better description of
the physics. One can imagine some ad-hoc rules to obtain for example a
correct picture of the reconnection process. But the probably most
interesting (and difficult) study is about the incorporation of the
non-linear dynamics in a more realistic way than the simple on-off
mechanism of CA models, but without going directly to the MHD
equations. From a pure observational point of view it seems crucial to
have as precise as possible an estimate of the possible biases to
determine the effective value of the power-law index $\zeta$ of the
energy distribution. Indeed the confirmation of the sub-critical value
of $\zeta$ could be a serious challenge to Parker's hypothesis of
coronal heating by a swarm of nanoflares.


\begin{acknowledgements}
  The authors acknowledge partial financial support from the PNST
  (Programme National Soleil--Terre) program of INSU (CNRS).
  \changed{M.\ Velli and G.\ Einaudi acknowledge support from MIUR
    contract MM02242342\_002.}  E.~Buchlin thanks the Scuola Normale
  Superiore of Pisa, Italy, for support and accommodation.
  \changed{They thank Dr.\ M.~Georgoulis for providing the bottom
    panel of figure \ref{fig:typfields}. They appreciate the useful
    criticism of the anonymous referee.}

\end{acknowledgements}


\begin{thebibliography}{Aletti et al.(2000)}
\bibitem[Aletti et al.(2000)]{Aletti00}
Aletti, V., Velli, M., Bocchialini, K., Einaudi, G., Georgoulis, M., 
\& Vial, J.-C. 2000, \apj, 544, 550

\bibitem[Aletti(2001)]{Aletti01}
Aletti, V. 2001,
PhD Thesis (University of Paris-XI)

\bibitem[Aschwanden et al.(2000)]{Asch00b}
Aschwanden, M.J., Tarbell, T.D., Nightingale, R.W., Schrijver, C.J.,
Title, A., Kankelborg, C.C., Martens, P., \& Warren, H.P. 2000,
\apj, 535, 1047 

\bibitem[Aschwanden \& Charbonneau(2002)]{Asch02}
Aschwanden, M.J., \& Charbonneau, P. 2002, \apj, 566, L59

\bibitem[Bak et al.(1987)]{Bak87} 
Bak, P., Tang, C., \& Wiesenfeld, K. 1987, Phys. Rev. Lett., 59, 381

\bibitem[Benz \& Krucker(2002)]{Benz}
Benz, A.O., \& Krucker, S. 2002, \apj, 568, 413

\bibitem[Berghmans et al.(1998)]{Berghmans}
Berghmans, D., Clette, F., \& Moses, D. 1998, A\&A, 336, 1039

\bibitem[Boffetta et al.(1999)]{Boffetta}
Boffetta, G., Carbone, V., Giuliani, P., Veltri, P., \& Vulpiani, A.
1999, Phys. Rev. Lett., 83, 4662

\bibitem[Carlson \& Langer(1989)]{carlson}
Carlson, J.M., \& Langer, J.S. 1989, Phys. Rev. A, 40, 6470 

\bibitem[Charbonneau et al.(2001)]{Charbon}
Charbonneau, P., McIntosh, S.W., Liu, H.-L., \& Bogdan, T.J. 2001, 
Solar Phys., 203, 321

\bibitem[Chou et al.(1991)]{chou}
Chou, D.-Y., LaBonte, B.J., Braun, D.C., \& Duvall Jr., T.L. 1991, 
\apj, 372, 314 

\bibitem[Christensen \& Olami(1992)]{chris92}
Christensen, K., \& Olami, Z. 1992, 
J. Geophys. Res., 97, 8729

\bibitem[Crosby, Aschwanden \& Dennis(1993)]{Crosby93} 
Crosby, N.B., Aschwanden, M.J., \& Dennis, B.R. 
1993, Solar Phys., 143, 275

\bibitem[Dennis(1985)]{Dennis85}
Dennis, B.R. 1985, Sol. Phys., 100, 465

\bibitem[Dmitruk et al.(1998)]{dmitruk98}
Dmitruk, P., G\'omez, D.O., \& DeLuca, E.E. 1998, \apj, 505, 974

\bibitem[Einaudi et al.(1996)]{Einaudi96}
Einaudi, G., Velli, M., Politano, H., \& Pouquet, A. 1996, 
\apj, 457, L113 

\bibitem[Einaudi \& Velli(1999)]{einaudi}
Einaudi, G., \& Velli, M. 1999, Phys. Plasmas, 6, 4146 

\bibitem[Espagnet et al.(1993)]{espagnet}
Espagnet, O., Muller, R., Roudier, Th., \& Mein, N. 1993, 
A\&A, 271, 589 

\bibitem[Galsgaard(1996)]{galsgaard}
Galsgaard, K. 1996, A\&A, 315, 312 

\bibitem[Galsgaard \& Nordlund(1996)]{GalsgaardN}
Galsgaard, K., \& Nordlund, A. 1996, J. Geophys. Res., 101, 13445

\bibitem[Galtier \& Pouquet(1998)]{Gal98}
Galtier, S., \& Pouquet, A. 1998, Solar Phys., 179, 141

\bibitem[Galtier(1999)]{Gal99}
Galtier, S. 1999, \apj, 521, 483 

\bibitem[Galtier(2001)]{Gal01}
Galtier, S. 2001, Solar Phys., 201, 133

\bibitem[Georgoulis \& Vlahos(1998)]{George98a}
Georgoulis, M.K., \& Vlahos, L. 1998, A\&A, 336, 721

\bibitem[Georgoulis et al.(1998)]{George98b}
Georgoulis, M.K., Velli, M., \& Einaudi, G. 1998, \apj, 497, 957

\bibitem[Hendrix \& Van Hoven(1996)]{Hendrix96}
Hendrix, D.L., \& Van Hoven, G. 1996, \apj, 467, 887

\bibitem[Hudson(1991)]{Hudson91}
Hudson, H.S. 1991, Solar Phys., 133, 357

\bibitem[Hwa \& Kardar(1992)]{hwa}
Hwa, T., \& Kardar, M. 1992, Phys. Rev. A, 45, 7002 

\bibitem[Isliker, Anastasiadis \& Vlahos(2000)]{Isliker00}
Isliker, H., Anastasiadis, A., \& Vlahos, L. 2000, A\&A, 363, 1134

\bibitem[Isliker, Anastasiadis \& Vlahos(2001)]{Isliker01}
Isliker, H., Anastasiadis, A., \& Vlahos, L. 2001, A\&A, 377, 1068 

\bibitem[Kadanoff et al.(1989)]{Kadanoff89}
Kadanoff, L.P., Nagel, S.R., Wu, L., \& Zhou, S. 1989, 
Phys. Rev. A, 39, 6524

\bibitem[Krasnoselskikh et al.(2001)]{krasno}
Krasnoselskikh, V., Podladchikova, O., Lefebvre, B., \& Vilmer, N.
2002, A\&A, 382, 699

\bibitem[Krucker \& Benz(1998)]{Krucker98}
Krucker, S., \& Benz, A. 1998, \apj, 501, L213

\bibitem[Lejeune \& Perdang(1996)]{lejeune}
Lejeune, A., \& Perdang, J. 1996, A\&ASS, 119, 249 

\bibitem[Lepreti, Carbone \& Veltri(2001)]{Lepreti}
Lepreti, F., Carbone, V. \& Veltri, P. 2001, \apj, 555, L133

\bibitem[Longcope \& Sudan(1994)]{Long94}
Longcope, D.W., \& Sudan, R.N. 1994, \apj, 437, 491

\bibitem[Lu \& Hamilton(1991)]{Lu91}
Lu, E.T., \& Hamilton, R.J. 1991, \apj, 380, L89

\bibitem[Lu et al.(1993)]{Lu93}
Lu, E.T., Hamilton, R.J., Mc Tiernan , J.M., \& Bromund, K.R. 1993, 
\apj, 412, 841

\bibitem[Lu(1995)]{Lu95}
Lu, E.T. 1995, Phys. Rev. Lett., 74, 2511 

\bibitem[MacKinnon \& Macpherson(1997)]{mac2}
MacKinnon, A.L., \& Macpherson, K.P. 1997, A\&A, 326, 1228 

\bibitem[Parker(1988)]{parker88}
Parker, E.N. 1988, \apj, 330, 474

\bibitem[Parnel \& Jupp(2000)]{parnell00}
Parnell, C.E. \& Jupp, P.E. 2000, \apj, 529, 554

\bibitem[Pearce, Rowe \& Yeung(1993)]{Pearce93}
Pearce, G., Rowe, A.K., \& Yeung, J. 1993, Ap\&SS, 208, 99

\bibitem[Podladchikova(2002)]{Podla02}
Podladchikova, O. 2002,
PhD Thesis (University of Orl\'eans, University of Kiev)



\bibitem[Roudier \& Muller(1986)]{roudier}
Roudier, Th., \& Muller, R. 1986, Solar Phys., 107, 11 

\bibitem[Sornette(2000)]{sornette}
Sornette, D. 2000, Critical phenomena in natural sciences
(Berlin\,: Springer-Verlag)

\bibitem[Strauss(1976)]{strauss}
Strauss, H.R. 1976, Phys. Fluids, 19, 134 

\bibitem[Takalo et al.(1999)]{takalo}
Takalo, J., Timonen, J., Klimas, A., Valdivia, J., \& Vassiliadis, D.
1999, Geophys. Res. Lett., 26, 1813 

\bibitem[Vassiliadis et al.(1998)]{vassiliadis}
Vassiliadis, D., Anastasiadis, A., Georgoulis, M., \& Vlahos, L. 
1998, \apj, 509, L53 

\bibitem[Vlahos et al.(1995)]{Vlahos95}
Vlahos, L., Georgoulis, M., Kluiving, R., \& Paschos, P. 1995, 
A\&A, 299, 897

\bibitem[Walsh, Bell \& Hood(1995)]{HOO}
Walsh, R.W., Bell, G.E., \& Hood, A.W. 1995, Solar Phys., 161, 83

\bibitem[Walsh \&  Galtier(2000)]{WG00} 
Walsh, R.W., \& Galtier, S. 2000, Solar Phys., 197, 57 

\bibitem[Wheatland, Sturrock \& McTiernan(1998)]{Wheatland}
Wheatland, M.S., Sturrock, P.A., \& McTiernan, J.M. 1998,
\apj, 509, 448

\bibitem[Wheatland(2000)]{Wheatland00}
Wheatland, M.S. 2000, \apj, 536, L109

\bibitem[Zirker(1993)]{zirker}
Zirker, J.B. 1993, Solar Phys., 148, 43
\end{thebibliography}
\end{document}